\documentclass[preprint, superscriptaddress, showpacs,preprintnumbers,amsmath,amssymb]{revtex4}
\usepackage{graphicx}

\begin{document}

\thispagestyle{empty}

\title{Constraints on non-Newtonian gravity and light elementary
particles from measurements of the Casimir force by means of
dynamic AFM}
\author{
G.~L.~Klimchitskaya,${}^1$ U.~Mohideen,${}^2$
and V.~M.~Mostepanenko}
\affiliation{Central Astronomical Observatory
at Pulkovo of the Russian Academy of Sciences,
St.Petersburg, 196140, Russia \\
and\\
Institute for Theoretical
Physics, Leipzig University, Postfach 100920,
D-04009, Leipzig, Germany\\
${}^{\it 2}$Department of Physics and
Astronomy, University of California, Riverside, California 92521,
USA}

\begin{abstract}
We derive constraints on corrections to Newtonian gravity
of the Yukawa type and light elementary particles
from two recently performed measurements of the gradient of the
Casimir force. In the first measurement the configuration of two
Au surfaces has been used, whereas in the second a nonmagnetic
metal Au interacted with a magnetic metal Ni. In these
configurations one arrives at
different, respectively, similar theoretical predictions for the
Casimir force when the competing theoretical approaches are
employed. Nevertheless, we demonstrate that the constraints
following from both experiments are in mutual agreement and in line
with constraints obtained from earlier measurements.
This confirms the reliability of constraints on non-Newtonian
gravity obtained from measurements of the Casimir force.
\end{abstract}

\maketitle
\section{Introduction}

Deviations from Newton's gravitational law are presently discussed
in many different aspects in connection with various extensions of
the Standard Model and the problem of dark matter.
The Yukawa-type corrections to Newtonian gravity in the wide
interaction range from nanometers to $10^{15}\,$cm are predicted
\cite{1,2} by the extra-dimensional models with a low-energy
compactification scale of the order of 1\,TeV \cite{3,4,5}.
The modified Newtonian dynamics is discussed \cite{6} as an
alternative to the proposed existence of dark matter which
should comprise about 23\% of the Universe mass.
On the other hand, if it is granted that the most realistic
according to astrophysics, cold, dark matter exists in nature,
the question arises as to what are its constituents.
One of the answers to this question is that the cold dark
matter consists of non-thermally produced axions \cite{8}.
The exchange of such type of particles between atoms of two
macrobodies generates an effective Yukawa-type correction
to Newton's gravitational law with an interaction range
$\lambda$ varying from 1\,{\AA}  to hundreds of meters or even
larger depending on the mass of a particle
$m=\hbar/(\lambda c)$.

Constraints on corrections to Newtonian gravity are under
investigation for many years \cite{9}.
Strong constraints within the interaction range from millimeters
to meters were obtained from the Cavendish- and E\"{o}tvos-type
experiments \cite{10,11}. In the submillimeter interaction range
stronger constraints on the Yukawa-type corrections were
currently derived from Cavendish-type experiments
\cite{12,13,14}.
The strength of these constraints, however, decreases with
decreasing $\lambda$. For $\lambda$ smaller than a few micrometers
the strongest constraints on the Yukawa-type corrections to
Newtonian gravity follow from measurements of the Casimir force
\cite{15,16} which becomes the dominant background force in place
of gravitation at sufficiently short separations between the test
bodies (see review \cite{15a}).
Measurements of the lateral Casimir force between
sinusoidally corrugated surfaces of a sphere and a plate by
means of an atomic force microscope (AFM) \cite{17,18} and the
gradient of the Casimir force between smooth surfaces of a sphere
and a plate by means of a micromachined oscillator \cite{19,20}
have already resulted \cite{21} in up to several orders of
magnitude stronger constraints.

Constraints on the parameters of non-Newtonian gravity are
commonly obtained from the measure of agreement between the
experimental data for the Casimir force and theoretical
predictions based on the Lifshitz theory \cite{15}.
It is necessary to stress, however, that theory-experiment
comparison in the field of the Casimir effect has led to
unexpected results which are yet to be fully explained.
Specifically, it was found that for metallic test bodies the
theoretical predictions are in agreement with the experimental
data if the relaxation properties of conduction electrons are
omitted \cite{15,16,19,20,22,23,24,25}.
An experiment \cite{26}, showing agreement between the data and theory
with the relaxation properties included, measured the sum of the
Casimir force and up to an order of magnitude larger residual
electric force supposedly originating from large patches. It was
not possible to independently measure this residual electric force.
Because of this, the model description for it has been used
containing two fitting parameters. The values of fitting parameters
were determined from the fit between the experimental data and
theory. As was pointed out in the literature \cite{27,28}, at short
separations Ref.~\cite{26} neglects the role of surface imperfections of
a spherical lens with centimeter-size radius of curvature. It was
also shown \cite{29,30} that even with account of patch potentials, at
large separations the experimental data of Ref.~\cite{26} for the total
force are in better agreement with theory neglecting the
contribution of charge carriers. Note that the issue concerning the
role of patch potentials has yet not been finally solved. The
experimental data of Refs.~\cite{22,23,24,25} are consistent with theory of
Ref.~\cite{30a},  predicting negligibly small
contribution of patches in the configurations of these experiments.
For an alternative theory of patches see Ref.~\cite{30b} and its
discussion in Ref.~\cite{24}.

For dielectric test bodies the theoretical predictions were
found in agreement with the data only if the contribution of free
charge carriers is omitted \cite{31,32,33,34,35,36}.
Keeping in mind that many of the experiments mentioned above were
used, first, to make a selection between different theoretical
approaches and, second, to constrain corrections to Newtonian
gravity from the measure of agreement between the data and the
predictions following from a selected approach, the constraints obtained
were sometimes claimed to be of dubious merit \cite{37}.
This claim is questionable \cite{38} because the difference
between the excluded and confirmed theoretical approaches to
the Casimir force cannot be modeled by the correction to
Newtonian gravity of Yukawa type.
It would be desirable, however, to have an
independent confirmation of the previously obtained constraints
that is not connected with a selection between different
competing models of the Casimir force.

In this paper we derive constraints on non-Newtonian gravity and
on the parameters of possible constituents of dark matter
following from two recent measurements
of the gradient of the Casimir force by means of dynamic AFM
operated in the frequency-shift technique \cite{24,25,39}.
The first of these experiments \cite{24,25} deals with a hollow,
Au-coated glass sphere oscillating in close proximity to an
Au-coated sapphire plate. In some sense the experiment
\cite{24,25}
is similar to experiments \cite{19,20,22,23}, but it is performed
using quite different laboratory setup (an AFM instead of a
micromachined oscillator).
The experimental data were compared with different theoretical
approaches to the Casimir force and again were found to be in
favor of the approach with the relaxation properties of
conduction electrons omitted \cite{24,25}. Here we obtain
constraints on non-Newtonian gravity following from this
experiment and demonstrate that they are in agreement with those
obtained in Refs.~\cite{19,20} (but slightly weaker in the same
proportion as the ratio of experimental errors in both
experiments).

In the second experiment considered in this paper the gradient
of the Casimir force acting between an Au-coated hollow glass
sphere and a Si plate coated with a ferromagnetic metal Ni was
measured by means of a dynamic AFM \cite{39}.
This configuration is of outstanding interest for constraining
corrections to Newtonian gravity because, as was shown in
Refs.~\cite{40,41}, within the range of experimental separations
one obtains almost coincident theoretical Casimir forces,
irrespective of whether the relaxation properties are included or
omitted. Thus, we arrive at constraints on the parameters of
non-Newtonian gravity which are independent of a selection between
different theoretical approaches to the Casimir force.
These constraints turn out to be slightly stronger than those
obtained from the AFM experiment with two Au surfaces, because
the magnetic experiment extends to smaller separation distances.
At the same time, the constraints obtained here from the magnetic
experiment are slightly weaker than those obtained previously
in Refs.~\cite{19,20} due to relatively lower precision of
AFM measurements, as compared with measurements performed by
means of a micromachined oscillator. Thus, the strongest
constraints on non-Newtonian gravity derived in Refs.~\cite{19,20}
receive further independent substantiation.

The paper is organized as follows. In Sec.~II we obtain constraints
on non-Newtonian gravity following from the dynamic AFM
experiment with two Au surfaces. In Sec.~III the same is done
using the measurement data of the AFM experiment with ferromagnetic
plate. Section~IV contains our conclusions and discussion.

\section{Constraints from the dynamic AFM experiment with
two Au test bodies}

The experiment \cite{24,25} considered in this section is in some
analogy to earlier performed experiments of
Refs.~\cite{19,20,22,23}, but it uses a dynamic AFM instead of a
micromechanical torsional oscillator as a measurement device.
The gradient of the Casimir force was measured between an
Au-coated hollow glass microsphere, attached to the cantilever
of an AFM, and an Au-coated sapphire plate within the range of
separations from 235 to 500\,nm. The thickness and density of
the glass spherical envelope were $\Delta^{\!(s)}=5\,\mu$m and
$\rho_{s}=2.5\times 10^{3}\,\mbox{kg/m}^3$.
For technological purposes the glass sphere was coated first with
a layer of Al having a thickness $\Delta_{\rm Al}^{\!(s)}=20\,$nm
and density $\rho_{\rm Al}=2.7\times 10^{3}\,\mbox{kg/m}^3$ and
then with a layer of Au having the thickness
$\Delta_{\rm Au}^{\!(s)}=280\,$nm and density
$\rho_{\rm Au}=19.28\times 10^{3}\,\mbox{kg/m}^3$.
The external radius of a coated sphere was measured to be
$R=41.3\,\mu$m \cite{24,25}.
The sapphire plate of density
$\rho_{p}=4.1\times 10^{3}\,\mbox{kg/m}^3$,
which can be considered as infinitely thick, was coated with
an Al layer of thickness
$\Delta_{\rm Al}^{\!(p)}=\Delta_{\rm Al}^{\!(s)}$ and
with an external Au layer of thickness
$\Delta_{\rm Au}^{\!(p)}=\Delta_{\rm Au}^{\!(s)}$.

The Casimir interaction between the sphere and the plate measured
in the experiment \cite{24,25} coexists with Newtonian
gravitation and possible corrections to it. For calculations of
the Casimir force both test bodies can be considered as made
of bulk Au \cite{15}. The gravitational force and corrections to
it should be calculated, however, taking into account the layer
structure of the sphere and plate. We admit that the
corrections
to  Newtonian gravitation has the Yukawa form, so that the
additional interaction to the
Casimir interaction between the two point-like masses
$m_1$ and $m_2$ can be described by the potential
\begin{equation}
V(r)=V_N(r)+V_{\rm Yu}(r)=-\frac{Gm_1m_2}{r}\left(
1+\alpha e^{-r/\lambda}\right).
\label{eq1}
\end{equation}
\noindent
Here, $r$ is the separation distance between the masses,
$G$ is the Newtonian
gravitational constant, $\alpha$ and $\lambda$ are
the strength and interaction range of the Yukawa interaction.
It can be easily seen \cite{42} that in the experimental
configurations under consideration the Newtonian gravitational
force is much smaller than the error in the measurement of the
Casimir force and can be neglected. The total Yukawa interaction
energy is obtained by the integration over the volumes of a
sphere $V_s$ and a plate $V_p$:
\begin{equation}
V_{\rm Yu}(a)=-G\alpha\int_{V_p}d^3r_1
\rho_p(\mbox{\boldmath$r$}_1)\int_{V_s}d^3r_2
\rho_s(\mbox{\boldmath$r$}_2)
\frac{e^{-|{\scriptsize{\mbox{\boldmath$r$}_1-
\mbox{\boldmath$r$}_2}}|/\lambda}}{|\mbox{\boldmath$r$}_1-
\mbox{\boldmath$r$}_2|},
\label{eq2}
\end{equation}
\noindent
where $\rho_p(\mbox{\boldmath$r$}_1)$ and
$\rho_s(\mbox{\boldmath$r$}_2)$ are the respective mass densities
and $a$ is the closest separation between the bodies.
Then the Yukawa force and its gradient are given by
\begin{equation}
F_{\rm Yu}(a)=-\frac{\partial V_{\rm Yu}(a)}{\partial a},
\qquad
\frac{\partial F_{\rm Yu}(a)}{\partial a}
=-\frac{\partial^2 V_{\rm Yu}(a)}{\partial a^2}.
\label{eq3}
\end{equation}

After performing the integration in Eq.~(\ref{eq2}) with account
of the layer structure of a sphere and a plate, one obtains from
Eq.~(\ref{eq3}) \cite{43}
\begin{equation}
\frac{\partial F_{\rm Yu}(a)}{\partial a}=
-4\pi^2G\alpha\lambda^2e^{-a/\lambda}X^{(s)}(\lambda)X^{(p)}(\lambda),
\label{eq4}
\end{equation}
\noindent
where the following notations are introduced:
\begin{eqnarray}
&&
X^{(p)}(\lambda)=\rho_{\rm Au}-(\rho_{\rm Au}-\rho_{\rm Al})
e^{-\Delta_{\rm Au}^{\!(p)}/\lambda}
\nonumber \\
&&~~~~
-(\rho_{\rm Al}-\rho_{p})
e^{-(\Delta_{\rm Au}^{\!(p)}+\Delta_{\rm Al}^{\!(p)})/\lambda},
\nonumber \\
&&
X^{(s)}(\lambda)=\rho_{\rm Au}\Phi(R,\lambda)-
(\rho_{\rm Au}-\rho_{\rm Al})\Phi(R-\Delta_{\rm Au}^{\!(s)},\lambda)
e^{-\Delta_{\rm Au}^{\!(s)}/\lambda}
\nonumber \\
&&~~
-(\rho_{\rm Al}-\rho_{s})\Phi(R-\Delta_{\rm Au}^{\!(s)}-\Delta_{\rm Al}^{\!(s)},\lambda)
e^{-(\Delta_{\rm Au}^{\!(s)}+\Delta_{\rm Al}^{\!(s)})/\lambda}
\nonumber \\
&&~~
-\rho_{s}\Phi(R-\Delta_{\rm Au}^{\!(s)}-\Delta_{\rm Al}^{\!(s)}-\Delta^{\!(s)},\lambda)
e^{-(\Delta_{\rm Au}^{\!(s)}+
\Delta_{\rm Al}^{\!(s)}+\Delta^{\!(s)})/\lambda},
\nonumber \\
&&
\Phi(r,\lambda)=r-\lambda+(r+\lambda)e^{-2r/\lambda}.
\label{eq5}
\end{eqnarray}

The experimental data for the gradient of the Casimir force were
found \cite{24,25} to exclude the theoretical
approach using the Lifshitz theory
and the dielectric permittivity obtained from the tabulated
optical data of Au extrapolated to zero-frequency by means of the
Drude model. The same force data turned out to be consistent with
theory when the simple plasma model is used for the extrapolation
to zero frequency within the limits of experimental errors
$\Delta_{F^{\prime}}(a)$ in the measurement of the force gradient
which were determined at the 67\% confidence level \cite{24,25}.
In the limits of these errors no interactions of Yukawa type were
observed. Thus, the constraints on the parameters of Yukawa
interaction $\alpha$ and $\lambda$ can be obtained from
the inequality
\begin{equation}
\left|\frac{\partial F_{\rm Yu}(a)}{\partial a}\right|
\leq\Delta_{F^{\prime}}(a).
\label{eq6}
\end{equation}

Equations (\ref{eq4}) and (\ref{eq5}) were substituted in
inequality (\ref{eq6}). It was shown that the strongest
constraints
follow at the shortest separation $a=235\,$nm. In
the measurement scheme with applied compensating voltage
$\Delta_{F^{\prime}}(a)=0.50\,\mu$N/m \cite{24}.
The resulting constraints are shown as line~1 in
Fig.~1, where the region of $(\lambda,\alpha)$ above
the line is prohibited and below the line is allowed by the
results of this experiment. The interaction region from
$\lambda_{\min}=20\,$nm to $\lambda_{\max}=3\,\mu$m
shown in Fig.~1 corresponds to the masses of a hypothetical
particle (axion, for instance) in the region from 66\,meV
to 9.9\,eV. This overlaps with the typical mass scale
from $m_a=1\,\mu$eV to $m_a=1\,$eV allowed for an axion
\cite{44} and includes part of the region allowed by the
cosmological bound for relic thermal axions
$m_a<0.42\,$eV \cite{45},
which are also considered in a cosmological context along
with non-thermally produced axions.
{}From the line 1 of Fig.~1 the allowed interaction
strength at $\lambda=\lambda_{\min}=20\,$nm is
$\alpha<5.9\times 10^{18}$. With increasing $\lambda$
(or, respectively, decreasing axion mass) the constraints
shown by line 1 become stronger. At
$\lambda=\lambda_{\max}=3\,\mu$m one obtains
$\alpha<3.6\times 10^{10}$.
In the next section the constraints of line 1 in Fig.~1
are compared with other constraints following from
measurements of the Casimir force.

In the end of this section we note that the constraints of
line 1 in Fig.~1 were obtained using the exact expressions
(\ref{eq4}), (\ref{eq5}) for the gradient of the Yukawa
force between a sphere and a plate \cite{43}.
Nearly the same results can be obtained in a more simple
way by calculating the gradient of the Yukawa force in the
framework of the proximity force approximation (PFA)
\begin{equation}
\frac{\partial F_{\rm Yu}(a)}{\partial a}=
-2\pi RP_{\rm Yu}(a),
\label{eq7}
\end{equation}
\noindent
where $P_{\rm Yu}(a)$ is the Yukawa pressure between two
parallel plates having the same layer structure as a plate and
a sphere in the experiment under consideration.
The approximate Eq.~(\ref{eq7}) is applicable under the
conditions
\begin{equation}
\frac{\lambda}{R}\ll 1, \qquad
\frac{\Delta_{\rm Au}^{\!(s)}+\Delta_{\rm Al}^{\!(s)}+
\Delta^{\!(s)}}{R}\ll 1,
\label{eq8}
\end{equation}
\noindent
which are satisfied in our case with a wide safety margin.
Using Eq.~(\ref{eq7}), one returns back to Eqs.~(\ref{eq4})
and (\ref{eq5}) where the function $\Phi(r,\lambda)$
with any argument $r$ is replaced with $R$ \cite{43,46}.
Now we compare the strength of constraints obtained
using the exact Yukawa force and the approximate one derived
using the PFA. Thus, at $\lambda$ equal to 1 and $3\,\mu$m
(recall that at larger $\lambda$ the accuracy of the PFA
is lower) the exact constraints are given by
$\alpha<2.19\times 10^{11}$ and
$\alpha<3.62\times 10^{10}$, respectively.
These should be compared with respective PFA results
$\alpha<2.17\times 10^{11}$ and
$\alpha<3.51\times 10^{10}$.
The comparison shows that the use of the PFA results in
only 0.9\% and 3.0\% relative errors at $\lambda=1$
and $3\,\mu$m, respectively.

\section{Constraints from the dynamic AFM experiment with
a magnetic plate}

The experiment described in the previous section was first
used for a selection between two different theoretical
approaches to the Casimir force. Then we have used the
measure of agreement between the selected approach and the
measurement data for obtaining constraints on non-Newtonian
gravity of the Yukawa type. This procedure is in fact well justified
because the difference between the two approaches to the
Casimir force cannot be modeled by the Yukawa interaction
with some $\lambda$ and $\alpha$. It would be interesting,
however, to independently verify the constraints obtained
in such a way by using the measurement data consistent with
the Lifshitz theory of the Casimir force without additional
conditions. A good opportunity for this is provided by the
recent measurement of the gradient of the Casimir force
between the hollow Au-coated glass sphere and Si plate
coated with ferromagnetic metal Ni \cite{39}.

The experiment on measuring the gradient of the Casimir force
between magnetic and nonmagnetic metals was performed by means
of a dynamic AFM in the configuration of a sphere and a plate.
The layer structure of a sphere was the same as in the
experiment of Sec.~II, but its external radius was equal to
$R=64.1\,\mu$m \cite{39}. The Si plate of density
$\rho_{\rm Si}=2.33\times 10^{3}\,\mbox{kg/m}^3$
was coated with a Ni layer of thickness
$\Delta_{\rm Ni}^{\!(p)}=154\,$nm and density
$\rho_{\rm Ni}=8.9\times 10^{3}\,\mbox{kg/m}^3$
with no additional intermediate layer.
As a result, the exact expression for the gradient of the
Yukawa force is again given by Eq.~(\ref{eq4}),
where $X^{(s)}(\lambda)$ is contained in Eq.~(\ref{eq5})
and $X^{(p)}(\lambda)$ takes a more simple form
\begin{equation}
X^{(p)}(\lambda)=\rho_{\rm Ni}-(\rho_{\rm Ni}-\rho_{\rm Si})
e^{-\Delta_{\rm Ni}^{\!(p)}/\lambda}.
\label{eq9}
\end{equation}

The measurement data for the gradient of the Casimir force over
the entire separation region from 220 to 500\,nm were found
consistent with the Lifshitz theory combined with the dielectric
permittivities of Au and Ni. The latter were obtained by using
the available optical properties of both metals extrapolated to
zero frequencies by means of either the Drude or the plasma
models. The important characteristic feature of this case is that,
by coincidence, over the region of experimental separations the
predictions of the Drude model approach almost coincide with the
predictions of the plasma model approach. Thus,
the experimental data for the gradient of the Casimir force
were found in agreement with the Lifshitz theory with no
additional selection process among the theoretical models.
This makes possible to use the magnetic experiment as an additional
independent test for the constraints on corrections to
Newtonian gravitation obtained previously.

The strongest constraints follow at the shortest separation
distance $a=220\,$nm. They are obtained from Eq.~(\ref{eq6}),
where now $\Delta_{F^{\prime}}(a)=0.79\,\mu$N/m (the larger
experimental error determined at the same 67\% confidence
level as in the experiment of Sec.~II is connected with the
larger value of the sphere radius), by the substitution of
Eq.~(\ref{eq4}), the quantity $X^{(s)}$ defined in
Eq.~(\ref{eq5}) and $X^{(p)}$ defined in Eq.~(\ref{eq9}).
The results are shown with line 2 in Fig.~1. As can be seen
in Fig.~1, both the lines 1 and 2 show qualitatively similar
constraints. In the region from $\lambda=\lambda_{\min}=20\,$nm
to $\lambda=\lambda_0=0.5\,\mu$m the constraints of line 2
(the experiment with magnetic metal) are slightly stronger
than the constraints of line 1 obtained from the experiment
with nonmagnetic metals. Furthermore, in the separation
region from $\lambda=\lambda_0=0.5\,\mu$m to
$\lambda=\lambda_{\max}=3\,\mu$m the constraints of the line 2
are slightly weaker than the constraints of the line 1.
This is explained by different minimum separations, where these
experiments have been performed, different sphere radii, and
different materials of the plate used.
There are, however, no qualitative differences that might be
connected with the fact that the experimental data of the
measurement with nonmagnetic bodies were used for making a
selection between the two different theoretical approaches to the
Casimir force, whereas the experimental data of the measurement
with a magnetic plate were not. {}From line 2,
at $\lambda=20\,$nm (the axion mass $m_a=9.9\,$eV) the interaction
strength of the Yukawa-type correction to Newtonian gravity is
constrained by $\alpha<4.2\times 10^{18}$ and
at $\lambda=3\,\mu$m (the axion mass $m_a=66\,$meV)
by $\alpha<5.6\times 10^{10}$.

For comparison purposes in Fig.~2 we again plot line 2
representing the constraints on the Yukawa interaction obtained
from the experiment using a magnetic plate \cite{39}
independently of any selection between the different theoretical
approaches to the Casimir force. In the same figure, the
constraints following from the dynamic determination of the
Casimir pressure by means of a micromachined oscillator
\cite{19,20} are shown by line 3, from the Casimir-less
experiment \cite{47}, where the Casimir force was compensated,
are shown by line 4, from measurement of the Casimir-Polder
force between rubidium atoms belonging to the Bose-Einstein
condensate and SiO${}_2$ plate \cite{33} are shown by line 5
\cite{21}, and from measurement of the Casimir force between
smooth Au-coated sphere and Si plate covered with nanoscale
trapezoidal corrugations \cite{48} are shown by line 6
\cite{49}.
The constraints of Ref.~\cite{49a} obtained from an experiment
\cite{26} are
not shown in Fig. 2 because they are stronger than those of lines 4
and 5 only at large interaction ranges
 ${\rm log}_{10}[\lambda\,\mbox{(m)}] > - 6.38$
and are not characterized by some definite confidence level.
Note that the constraints of lines 3 and 4 were
determined at a 95\% confidence level, and the constraints
of lines 5 and 6, as well as of line 2, were found at a
67\% confidence level.
{}From Fig.~2 it can be seen that although the constraints
of line 2 are not the strongest ones, they are quite
competitive within some interaction range and, as it is
independent of a selection process between different
theoretical approaches to the Casimir force,
they provide
additional support to the constraints obtained from
other experiments.

\section{Conclusions and discussion}

In the foregoing, we have obtained constraints on the Yukawa-type
corrections to Newton's gravitational law  and light elementary
particles which follow from the two recently performed
experiments on the Casimir force. In the first of these
experiments the gradient of the Casimir force between two Au
surfaces has been measured by means of a dynamic AFM \cite{24,25}.
This experiment, as well as the previous measurement performed
using another laboratory technique \cite{19,20}, was used for
both the selection between two competing theoretical approaches
to the Casimir force and for constraining corrections to
Newtonian gravity. In the second recently performed experiment
the dynamic AFM was used to measure the gradient of the Casimir
force between a nonmagnetic metal Au and a ferromagnetic metal
Ni \cite{39}. The unique feature of this experiment is that
in the region of experimental separations both competing
approaches to the theoretical description of the Casimir force lead
to nearly coincident results. For this experiment the unambiguous
theoretical prediction was found in agreement with the
experimental data and this fact has been used for obtaining
constraints on the corrections to Newtonian gravity.
Good agreement between the constraints obtained from both
recent experiments and from other experiments on measuring
the Casimir force performed earlier was demonstrated.
This allows to conclude that in spite of the widely known
problems in theory-experiment comparison discussed in the
literature, measurements of the Casimir force in laboratory
remain a reliable source of constraints on non-Newtonian
gravity of the Yukawa-type and light elementary particles within
the interation range from nanometers to micrometers.

\section*{Acknowledgments}
This work was supported by the DFG grant No.\ BO\ 1112/21--1
(G.L.K.\ and V.M.M.) and by the NSF Grant No.\ PHY0970161 (U.M.).
 G.L.K.\ and V.M.M.\ are grateful to the Institute for Theoretical
 Physics, Leipzig University for their kind hospitality.

\newpage
\begin{figure}[b]
\vspace*{-11cm}
\centerline{\hspace*{3cm}
\includegraphics{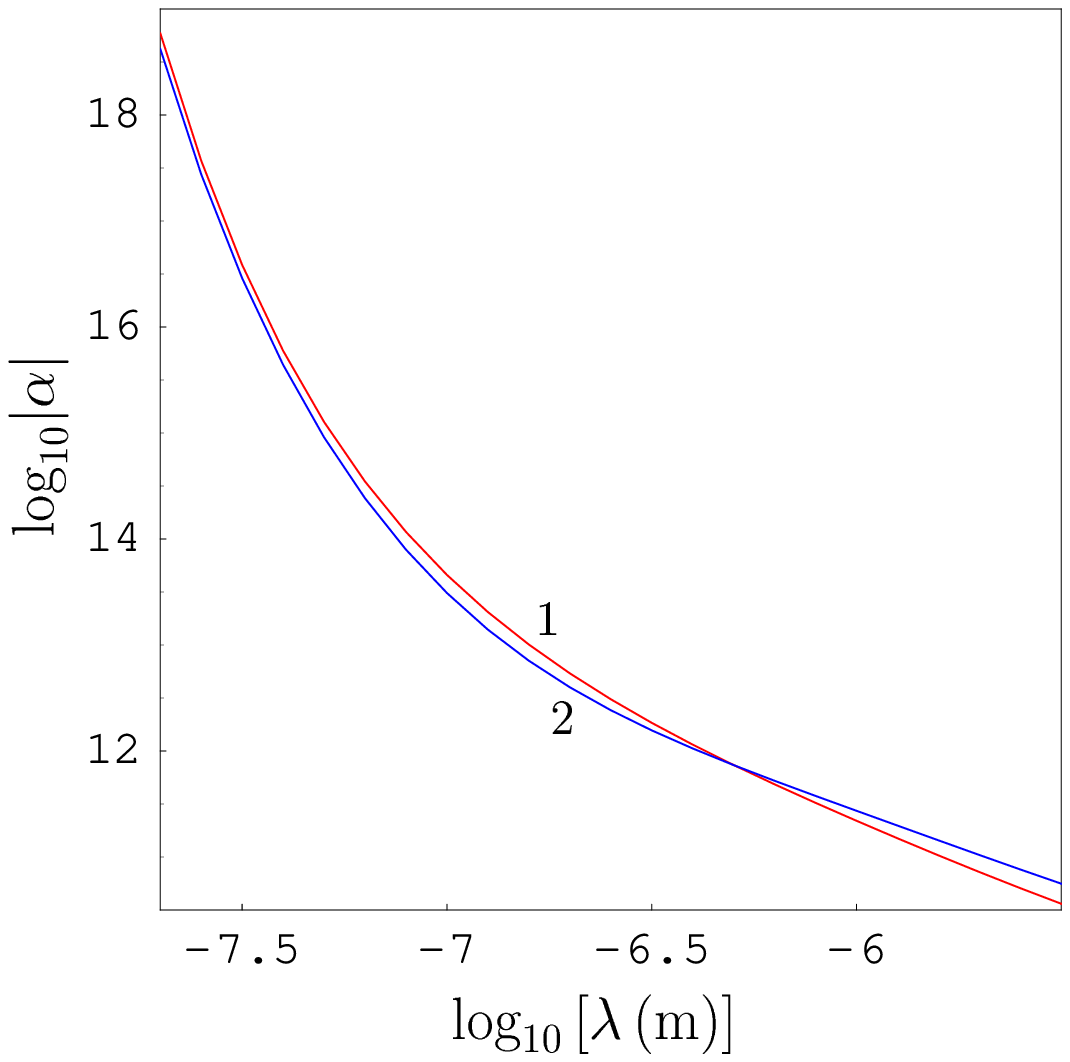}
}
\vspace*{-6cm}
\caption{(color online).
Constraints on the parameters of Yukawa-type correction to
Newton's gravitational law obtained from measurements of the
gradient of the Casimir force by means of a dynamic AFM in the
configuration of an Au-coated sphere and an Au-coated plate
(line 1) and in the
configuration of an Au-coated sphere and a magnetic
Ni-coated plate (line 2).
The regions of $(\lambda,\alpha)$ plane below each line are
allowed and above each line are prohibited.
}
\end{figure}
\begin{figure}[b]
\vspace*{-11cm}
\centerline{\hspace*{3cm}
\includegraphics{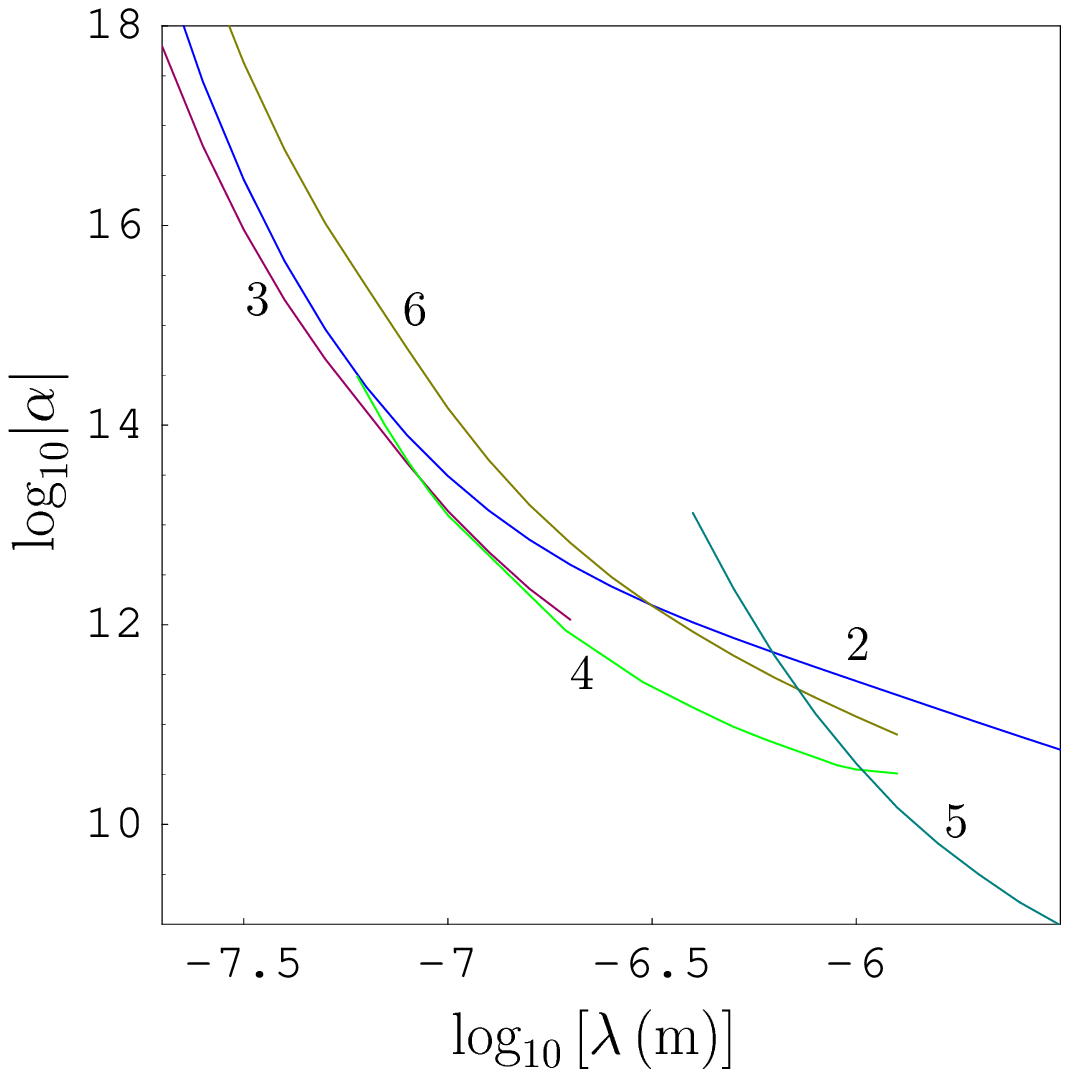}
}
\vspace*{-6cm}
\caption{(color online).
Constraints on the parameters of Yukawa-type correction to
Newton's gravitational law obtained from experiments
on the Casimir force with a magnetic plate performed by
means of a dynamic AFM \cite{39} (line 2) and with two Au
bodies performed by means of micromachined oscillator
\cite{19,20} (line 3), from the Casimir-less experiment
\cite{47} (line 4), from measurements \cite{33} of the
Casimir-Polder force \cite{21} (line 5), and from
measurements of the Casimir force between an Au sphere and
corrugated Si plate \cite{48,49} (line 6).
The regions of $(\lambda,\alpha)$ plane below each line are
allowed and above each line are prohibited.
}
\end{figure}

\begin{thebibliography}{99}
\bibitem{1}
E.~G.~Floratos and G.~K.~Leontaris,
Phys. Lett. B {\bf 465}, 95 (1999).
\bibitem{2}
A.~Kehagias and K.~Sfetsos,
Phys. Lett. B {\bf 472}, 39 (2000).
\bibitem{3}
I.~Antoniadis,
N.~Arkani-Hamed, S.~Dimopoulos, and G.~Dvali,
Phys. Lett. B {\bf 436}, 257 (1998).
\bibitem{4}
N.~Arkani-Hamed, S.~Dimopoulos, and G.~Dvali,
Phys. Lett. B {\bf 429}, 263 (1998).
\bibitem{5}
N.~Arkani-Hamed, S.~Dimopoulos, and G.~Dvali,
Phys. Rev. D {\bf 59}, 086004 (1999).
\bibitem{6}
J.~D.~Bekenstein,
Contemp. Phys. {\bf 47}, 387 (2006).
\bibitem{8}
R.~D.~Peccei and H.~R.~Quinn,
Phys. Rev. Lett. {\bf 38}, 1440 (1977).
\bibitem{9}
E.~Fischbach and C.~L.~Talmadge, {\it The Search for Non-Newtonian
Gravity} (Springer, New York, 1999).
\bibitem{10}
E.~G.~Adelberger, B.~R.~Heckel, C.~W.~Stubbs, and W.~F.~Rogers,
 Ann. Rev. Nucl. Part. Sci. {\bf 41}, 269 (1991).
\bibitem{11}
E.~G.~Adelberger, B.~R.~Heckel, and A.~E.~ Nelson,
 Ann. Rev. Nucl. Part. Sci. {\bf 53}, 77 (2003).
\bibitem{12}
S.~J.~Smullin, A.~A.~Geraci, D.~M.~Weld, J.~Chiaverini,
S.~Holmes, and A.~Kapitulnik,
Phys. Rev. D {\bf 72}, 122001 (2005).
\bibitem{13}
 A.~A.~Geraci, S.~J.~Smullin, D.~M.~Weld, J.~Chiaverini,
and A.~Kapitulnik,
{Phys. Rev. D} {\bf 78}, 022002 (2008).
\bibitem{14}
E.~G.~Adelberger, J.~H.~Gundlach, B.~R.~Heckel, S.~Hoedl,
and S.~Schlamminger,
Progr. Part. Nucl. Phys. {\bf 62}, 102 (2009).
\bibitem{15}
M.~Bordag, G.~L.~Klimchitskaya, U.\ Mohideen, and
V.\ M.\ Mostepanenko, {\it Advances in the Casimir Effect}
(Oxford University Press, Oxford, 2009).
\bibitem {16}
G.~L.~Klimchitskaya, U. Mohideen, and V.\ M.\ Mostepanenko,
Rev. Mod. Phys. {\bf 81}, 1827 (2009).
\bibitem{15a}
R.~Onofrio,
New J. Phys. {\bf 8}, 237 (2006).
\bibitem{17}
H.-C.\ Chiu,  G.~L.~Klimchitskaya, V.\ N.\ Marachevsky,
V.\ M.\ Mos\-te\-pa\-nen\-ko, and U.~Mohideen,
Phys. Rev. B {\bf 80}, 121402(R) (2009).
\bibitem{18}
H.-C.\ Chiu,  G.~L.~Klimchitskaya, V.\ N.\ Marachevsky,
V.\ M.\ Mos\-te\-pa\-nen\-ko, and U.~Mohideen,
Phys. Rev. B {\bf 81}, 115417 (2010).
\bibitem{19}
R.~S.~Decca, D.~L\'opez, E.~Fischbach, G.~L.~Klimchitskaya,
 D.~E.~Krause, and V.~M.~Mostepanenko,
Phys. Rev. D {\bf 75}, 077101 (2007).
\bibitem{20}
R.~S.~Decca, D.~L\'opez, E.~Fischbach, G.~L.~Klimchitskaya,
 D.~E.~Krause, and V.~M.~Mostepanenko,
Eur. Phys. J. C {\bf 51}, 963 (2007).
\bibitem{21}
V.~B.~Bezerra, G.~L.~Klimchitskaya,
 V.~M.~Mostepanenko, and C.~Romero,
Phys. Rev. D {\bf 81}, 055003 (2010).
\bibitem{22}
R.~S.~Decca,  E.~Fischbach, G.~L.~Klimchitskaya, D.~E.~Krause,
D.~L\'opez, and V.~M.~Mostepanenko,
Phys. Rev. D {\bf 68}, 116003 (2003).
\bibitem{23}
R.~S.~Decca, D.~L\'opez, E.~Fischbach, G.~L.~Klimchitskaya,
 D.~E.~Krause, and V.~M.~Mostepanenko,
 Ann. Phys. (N.Y.) {\bf 318}, 37 (2005).
\bibitem{24}
C.-C.~Chang, A.~A.~Banishev, R.~Castillo-Garza,
G.~L.~Klimchitskaya, V.\ M.\ Mostepanenko, and U.\ Mohideen,
Phys. Rev. B {\bf 85}, 165443 (2012).
\bibitem{25}
C.-C.~Chang, A.~A.~Banishev, R.~Castillo-Garza,
G.~L.~Klimchitskaya, V.\ M.\ Mostepanenko, and U.\ Mohideen,
Int. J. Mod. Phys.: Conf. Ser. {\bf 14}, 270 (2012).
\bibitem{26}
A.~O.~Sushkov, W.~J.~Kim, D.~A.~R.\ Dalvit,
and S.~K.~Lamoreaux,
{Nature Phys.} {\bf 7}, 230 (2011).
\bibitem{27}
V.~B.~Bezerra, G.~L.~Klimchitskaya, U.~Mohideen,
V.~M.~Mostepanenko, and C.~Romero,
{Phys. Rev. B} {\bf 83}, 075417 (2011).
\bibitem{28}
G.~L.~Klimchitskaya  and V.~M.~Mostepanenko,
Int. J. Mod. Phys. A {\bf 26}, 3944 (2011).
\bibitem{29}
G.~L.~Klimchitskaya, M.~Bordag,  E.~Fischbach, D.~E.~Krause,
and V.~M.~Mostepanenko,
Int. J. Mod. Phys. A {\bf 26}, 3918 (2011).
\bibitem{30}
G.~L.~Klimchitskaya, M.~Bordag,  and V.~M.~Mostepanenko,
Int. J. Mod. Phys. A {\bf 27}, 1260012 (2012).
\bibitem{30a}
C.~C.~Speake and C.~Trenkel,
Phys. Rev. Lett. {\bf 90}, 160403 (2003).
\bibitem{30b}
R.~O.~Behunin, F.~Intravaia, D.~A.~R.~Dalvit, P.~A.~Maia Neto,
and S.~Reynaud, Phys. Rev. A {\bf 85}, 012504 (2012).
\bibitem{31}
F.~Chen, G.~L.~Klimchitskaya, V.\ M.\ Mostepanenko, and U.\ Mohideen,
Optics Express  {\bf 15}, 4823 (2007).
\bibitem{32}
F.~Chen, G.~L.~Klimchitskaya, V.\ M.\ Mostepanenko, and U.\ Mohideen,
Phys. Rev. B  {\bf 76}, 035338 (2007).
\bibitem{33}
J.~M.~Obrecht, R.~J.~Wild, M.~Antezza, L.~P.~Pitaevskii,
S.~Stringari, and E.~A.~Cornell,
Phys. Rev. Lett. {\bf 98}, 063201 (2007).
\bibitem{34}
G.~L.~Klimchitskaya  and V.~M.~Mostepanenko,
J. Phys. A: Math. Theor. {\bf 41}, 312002 (2008).
\bibitem{35}
C.-C.~Chang, A.~A.~Banishev,
G.~L.~Klimchitskaya, V.\ M.\ Mostepanenko, and U.\ Mohideen,
Phys. Rev. Lett.  {\bf 107}, 090403 (2011).
\bibitem{36}
A.~A.~Banishev, C.-C.~Chang, R.~Castillo-Garza,
G.~L.~Klimchitskaya, V.\ M.\ Mostepanenko, and U.\ Mohideen,
Phys. Rev. B {\bf 85}, 045436 (2012).
\bibitem{37}
A.~Lambrecht, A.~Canaguier-Durand, R.~Gu\'{e}rout, and
S.~Reynaud,
In: {\it Casimir Physics}, Lecture Notes in Physics,
{\bf 834}, p.97,
eds. D.~A.~R.~Dalvit, P.~W.~Milonni, D.~C.~Roberts, and
F.~S.~S.~Rosa (Springer, Heidelberg, 2011).
\bibitem{38}
V.~M.~Mostepanenko, V.~B.~Bezerra, G.~L.~Klimchitskaya,
 and C.~Romero,
Int. J. Mod. Phys. A {\bf 27}, 1260015 (2012).
\bibitem{39}
A.~A.~Banishev, C.-C.~Chang,
G.~L.~Klimchitskaya, V.\ M.\ Mostepanenko, and U.\ Mohideen,
Phys. Rev. B {\bf 85}, 195422 (2012).
\bibitem{40}
B.~Geyer, G.~L.~Klimchitskaya, and V.~M.~Mostepanenko,
Phys. Rev. B  {\bf 81}, 104101 (2010).
\bibitem{41}
G.~L.~Klimchitskaya, B.~Geyer, and V.~M.~Mostepanenko,
 Int. J. Mod. Phys. A  {\bf 25}, 2293 (2010).
\bibitem{42}
M.~Bordag, B.~Geyer, G.~L.~Klimchitskaya,
and V.~M.~Mostepanenko,
{Phys. Rev. D}  {\bf 62}, 011701(R) (2000).
\bibitem{43}
R.~S.~Decca,  E.~Fischbach, G.~L.~Klimchitskaya,
 D.~E.~Krause, D.~L\'opez, and V.~M.~Mostepanenko,
Phys. Rev. D {\bf 79}, 124021 (2009).
\bibitem{44}
G.~G.~Raffelt, J.~Redondo, and N.~V.~Maira,
Phys. Rev. D {\bf 84}, 103008 (2011).
\bibitem{45}
A.~Melchiorri, O.~Mena, and A.~Slosar,
Phys. Rev. D {\bf 76}, 041303(R) (2007).
\bibitem{46}
E.~Fischbach, G.~L.~Klimchitskaya,
 D.~E.~Krause, and V.~M.~Mostepanenko,
Eur. Phys. J. C {\bf 68}, 223 (2010).
\bibitem{47}
R.~S.~Decca, D.~L\'opez, E.~Fischbach,
 D.~E.~Krause, and C.~R.~Jamell,
Phys. Rev. Lett. {\bf 94}, 240401 (2005).
\bibitem{48}
Y.~Bao, R.~Gu\'{e}rout, J.~Lussange, A.\ Lambrecht, R.\ A.\ Cirelli, F.\ Klemens,
W.\ M.\ Mansfield, C.\ S.\ Pai,
and H.~B.~Chan,
{Phys. Rev. Lett.} {\bf 105}, 250402 (2010).
\bibitem{49}
V.~B.~Bezerra, G.~L.~Klimchitskaya,
 V.~M.~Mostepanenko, and C.~Romero,
Phys. Rev. D {\bf 83}, 075004 (2011).
\bibitem{49a}
A.~O.~Sushkov, W.~J.~Kim, D.~A.~R.\ Dalvit,
and S.~K.~Lamoreaux,
Phys. Rev. Lett. {\bf 107}, 171101 (2011).
\end{thebibliography}
\end{document}